\documentclass[pre,10pt, a4paper,superscriptaddress,nofootinbib,twocolumn,longbibliography,showkeys]{revtex4-1}
\usepackage{graphicx}

\usepackage{placeins} 
\PassOptionsToPackage{hyphens}{url}
\usepackage{hyperref} 
\usepackage{bm}
\usepackage{amsmath} 
\usepackage[parfill]{parskip} 
\usepackage{amssymb,latexsym} 
\usepackage{amsthm}
\usepackage{amsmath}
\usepackage{mathtools}
\usepackage{setspace}
\usepackage{natbib}
\usepackage{url}
\usepackage{breakurl}
\usepackage{float}
\usepackage{xcolor}
\usepackage{cancel}
\usepackage{soul}
\usepackage{units}
\usepackage[export]{adjustbox}
\begin{document}
\title{\textit{Copepods} encounter rates\\ from a model of escape jump behaviour in turbulence}

\author{H. Ardeshiri}
\email{hamidreza.ardeshiri@polytech-lille.fr}
\affiliation{Univ. Lille, CNRS, FRE 3723, LML, Laboratoire de M\'ecanique de Lille, F 59000 Lille, France}
\affiliation{Univ. Lille, CNRS, Univ. Littoral Cote d'Opale, UMR 8187, LOG, Laboratoire d'Oc\'eanologie et de G\'eoscience, F 62930 Wimereux, France}
\author{F. G. Schmitt}
\author{S. Souissi}
\affiliation{Univ. Lille, CNRS, Univ. Littoral Cote d'Opale, UMR 8187, LOG, Laboratoire d'Oc\'eanologie et de G\'eoscience, F 62930 Wimereux, France}
\author{F. Toschi}
\affiliation{Department of Applied Physics and Department of Mathematics and Computer Science, Eindhoven University of Technology, 5600 MB, Eindhoven, The Netherlands}
\affiliation{Istituto per le Applicazioni del Calcolo CNR, Via dei Taurini 19, 00185 Rome, Italy}
\author{E. Calzavarini}
\affiliation{Univ. Lille, CNRS, FRE 3723, LML, Laboratoire de M\'ecanique de Lille, F 59000 Lille, France}





\begin{abstract}
A key ecological parameter for planktonic copepod studies is their encounter rates within the same population as well as with other species. The encounter rate is partly determined by copepod's swimming behaviour and is strongly influenced by turbulence of the surrounding environment. A distinctive feature of copepods' motility is their ability to perform quick displacements, often dubbed jumps, by means of powerful swimming strokes. Such a reaction has been associated to an escape behaviour from flow disturbances due to predators or other external signals. In the present study, we investigate the encounter rates of copepods from the same species in a developed turbulent flow with intensities comparable to those encountered in their natural habitat. This is done by means of a Lagrangian copepod (LC) model that mimics the jump escape reaction behaviour from localised high-deformation rate fluctuations in the turbulent flows. Our analysis shows that the encounter rate for copepods of typical perception radius of $\sim\eta$, where $\eta$ is the dissipative scale of turbulence, can be increased by a factor up to $\sim 10^2$ compared to the one experienced by passively transported fluid tracers of the same size. Furthermore, we address the effect of a minimal waiting time between consecutive jumps. It is shown that any encounter-rate enhancement is lost if such time goes beyond the dissipative time-scale of turbulence, $\tau_{\eta}$. Because typically in the ocean $\eta \sim 1\ mm$ and $\tau_{\eta} \sim 1\ s$, this provides stringent constraints on the turbulent-driven enhancement of contact-rate due to a purely mechanical induced escape reaction. The implications of our results in the context of ecology of copepods were discussed.
\end{abstract}
\maketitle

\setcounter{secnumdepth}{0}

\section{Introduction}
\label{Introduction}
Many biological processes are determined by individual-interaction or contacts between organisms. This is an essential aspect in plankton marine biology, because the encounter between individual organisms is vital for mating or for predation \cite{Hein-2013, Kiorboe-book, Menden-Deuer-2006}. Finding a suitable habitat (colonization) for marine organisms is also encounter dependent \cite{Wosniack-2014}. Organism size, morphology, motility and abundance can affect the encounter rates, \textit{i.e.}, the typical frequency at which individuals meet other organisms of the same or of different species. Additionally, the encounter rate is influenced by external environmental factors, such as hydrodynamic turbulence or hydrography. It is thus of great ecological importance to understand the combined biological and physical factors that can affect the encounter rates in oceanic flows. Here we review how the encounter rate estimation of plankton species has been evaluated in the past and explain why the proposed approaches lack crucial aspects of plankton dynamics.\\
The encounter rate of plankton has been studied extensively in the past \cite{Gerritsen-1977, Rothschild-1988, Evans-1989,MacKenzie-1994,Visser-1998, Kiorboe-1995}. Gerritsen and Strickler \cite{Gerritsen-1977} first introduced a model of plankton contact rate in a steady uniform flow. Despite its influential role, this model relied on an oversimplified description of the fluid flow environment. Natural habitats of plankton are rarely characterized by steady laminar flows, but are frequently turbulent and experience time and space dependent fluctuations. It is now commonly accepted that turbulence must have an influence on the dispersal, feeding and reproduction of plankton \cite{Pecseli-2012}. Furthermore, it is believed that both small and large eddies of turbulence can affect the encounter process \cite{Visser-1998}.\\
In the past some authors have regarded the small-scale turbulent processes as a homogenizing factor, so that their encounter rate models assumed the distribution of plankton to be homogeneous in space and time \cite{Kiorboe-1995, Sundby-1990, Davis-1991, MacKenzie-1991, Caparroy-1996, Dzierzbicka-Glowacka-2006a, Dzierzbicka-Glowacka-2006b}. However, in the fluid dynamics context it is now well known that turbulence can both increase spatial heterogeneity at small-scales (preferential concentration) and produce persistent clusters over time (this happens for instance for transported scalar fields such as temperature and for material particles carried by a flow) \cite{Frisch-book, Toschi-2009, Schmitt-book}.\\ 
Nonhomogeneous distribution of particles has been studied comprehensively; heavy particles (particles denser than the fluid) concentrate in low vorticity and high strain rate regions \cite{Toschi-2009,Maxey-1987, Squires-1991, Maxey-1987, Wang-1993, Fessler-1994, Zaichik-2006}, light particles are trapped by vortices in the flow \cite{Toschi-2009, Squires-1991, Zaichik-2006, Calzavarini-2008a, Calzavarini-2008b}, and particles without inertia (fluid passive tracers) are passively advected by the flow. In these cases the observed phenomenon of preferential concentration is controlled by the particles' Stokes number $St$ (which is the ratio of the aerodynamic response time of a particle over the turbulent characteristic time scale, $\tau_\eta$), and by the density contrast between the particle over the fluid density. 
The previous studies focused on non-living particles. But when dealing with living particles, a second important aspect of the problem is that the encounter rate of swimming organisms in a flow is also governed by biologically-driven processes \cite{Kiorboe-1995, Seuront-2001, Dzierzbicka-Glowacka-2006b}. Plankton, and in particular copepods, have specific swimming strategies which can be induced by external mechanical stimuli. Some studies in the past have assumed that physically-driven (turbulent transport) and biologically-driven (swimming) processes could be summed up linearly in the estimation of encounter rates \cite{Kiorboe-1995, Seuront-2001, Dzierzbicka-Glowacka-2006b}. This approach has been questioned on the basis of experimental evidence showing that the two contributions are entangled and most probably depend on each other. In other words, turbulence may induce changes in the behavioural swimming responses of microorganisms \cite{Jennifer-2012, Durham-2012}. Michalec et al., \cite{Michalec-2015a} provided an experimental evidence of the interaction between turbulence intensity and copepods swimming behaviour. We assume here that the theoretical model for the estimation of the collision-rate of material particles transported by a turbulent flow, developed by Sundaram and Collins, \cite{Sundaram-1977}, Wang et al. \cite{Wang-1998}, Reade and Collins, \cite{Reade-2000} and Collins and Keswani, \cite{Collins-2004} can be used also for encounter of micro-organisms, until the perception distance is reached. This formulation indeed allows us to take into account both the contribution to encounter rates associated with the spatial accumulation of organisms, due to the carrying flow, as well as the contribution associated to their velocity, which is mostly associated to their swimming behaviour. Despite many studies which have been performed on the effect of preferential concentration on coagulation of colloidal particles \cite{Wang-2000, Lian-2013, Falkovich-2002, Brunk-1998}, and of inertial particles \cite{Squires-1991, Monchaux-2010}, only few studies \cite{Squires-1995,Schmitt-2008} are available in the context of copepods ecology.\\
The goal of this work is to offer an estimate of the mutual encounter rate of copepods within the same species (no predator or other species have been introduced yet in the present model and all particles obey to the same laws), which builds on a previously proposed behavioural model of copepods in a flow referred to as the LC (Lagrangian Copepod) model \cite{Ardeshiri-2016}. The LC model was developed from an experimental data analysis input, in order to explore copepods' dynamics in developed turbulent flows. It was observed that jump escape reaction from spatiotemporal events characterized by high strain-rate results in copepods' nonhomogeneous spatial distribution in turbulent flows. In Ardeshiri et al., \cite{Ardeshiri-2016} only the description of these spatiotemporal patterns of particles was done. In this paper the objective is to estimate directly the encounter rate of copepods. Our hypothesis is that the encounter rate enhancement has some trend related to optimal concentration.
In order to test this hypothesis, the LC model in turbulence as in Ardeshiri et al., \cite{Ardeshiri-2016} is presented in a simpler mathematical form. As a further development to the model, an additional parameter, the waiting time between successive jumps in copepod, is introduced and its effects on their encounter rate are discussed. To the best of our knowledge, this is the first numerical simulation of swimming copepods behaviour in a realistic turbulent flows which under some specific conditions results in spatial clustering and enhanced encounter rates.\\

\section{METHOD}
\label{METHOD}
In this section we introduce the LC model which is adopted in the present study to quantify copepods' encounter rates in a turbulent flow. This model, which has been first presented in Ardeshiri et al., \cite{Ardeshiri-2016}, has the advantage to keep into account at the same time a representation of behavioural (jump escape reaction from intense turbulence) and hydrodynamical (flow advection) effects. In the present work, we use the same model and only consider a waiting time between successive jumps of copepods in order to estimate their potential effect on the encounter rate. Furthermore, the focus of the present work is more ecology-oriented and especially encounter rates are estimated.\\
The LC model is based on a Eulerian-Lagrangian modelling approach. This means that the fluid flow is obtained by solving the incompressible Navier-Stokes equations, by means of Direct Numerical Simulation (DNS), while the organisms position are treated in the same fashion as Lagrangian point-particles drifting by the flow. 

\subsection{Lagrangian copepod model}
\label{Lagrangian copepod model}
\begin{figure}
  \centering
             \includegraphics[clip, trim=0cm 4cm 0cm 2cm, scale = 0.4]{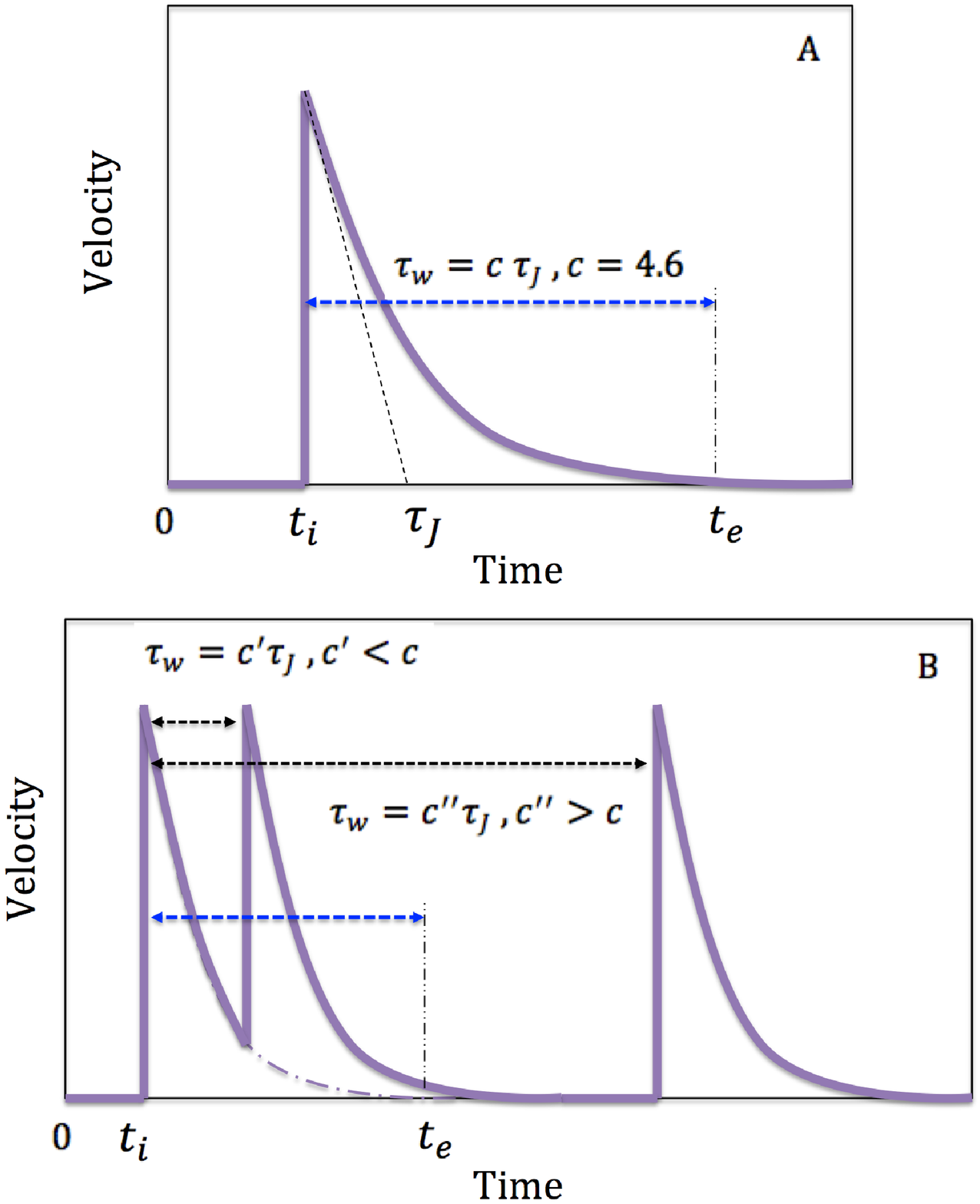}
  \caption{(A) Cartoon of the functional behaviour of the velocity during a single jump assuming the absence of the fluid velocity.
  (B) Same as before for two successive jumps at varying the duration of the time interval $\tau_w$ between them. Our reference value is here $\tau_w = t_e - t_i = c\ \tau_J$ with $c=4.6$. In the parametric study of section \textit{``Effect of waiting time between successive jumps"} we vary $\tau_w$ in the range $[2,1000] \tau_J$.}
  \label{jump}
\end{figure}

The Lagrangian model of copepods dynamics relies both on biological and hydrodynamical assumptions. i) A first assumption is that copepods escape reactions are triggered by a well defined mechanical signal, the strain rate $\dot\gamma$, and that a jump begins whenever this signal exceeds a fixed threshold value, denoted with $\dot\gamma_T$. 
The rate of strain intensity is a quantity which depends on the velocity gradients, it is defined as:
\begin{equation}
\dot\gamma = \sqrt{ \Sigma_{i=1}^3 \Sigma_{j=1}^{3} 1/2 (\partial_i u_j + \partial_j u_i)^2 }
\label{gammadot}
\end{equation}
where $u_i$ is the fluid velocity field in three dimensions ($i=1,2, 3$) ($u_j$  is the same quantity as $u_i$ referring to other components of the velocity field corresponding to its index ($j=1,2, 3$)). We note that the strain rate include both normal strain, $\partial_i u_i$, and shear strain, $\partial_j u_i \ (i\neq j)$. In turbulent flows, due to local isotropy, the latter term is the dominant one \cite{Hinze-book} for this reason we will also refer to $\dot{\gamma}$ as to the shear rate.
We also note that the proposed first assumption makes several simplifications. In particular it neglects any other copepod swimming activity induced by light, food, or chemistry (\textit{e.g.,} pheromones). Moreover, it neglects the fact that copepod may want to avoid too calm regions of the flow, in other words regions where $\dot\gamma$ could be below a certain threshold (a behaviour that has also been reported in the literature \cite{Saiz-1995}).\\
ii) A second assumption is that copepods response is always in the form of an escape reaction, independently of the intensity of the external mechanical disturbance. Experimental evidence suggests that the velocity tracks of copepods performing jumps show always a very sharp velocity increase, followed by an approximately exponential decay in amplitude. This has been highlighted in high-speed recordings for \textit{Eurytemora affinis} and \textit{Acartia tonsa} in still water experiments \cite{Ardeshiri-2016, Ardeshiri-thesis-2016}. Such a functional dependence for the velocity can be associated to the effect of an impulsive force due to a stroke (or a burst of strokes) that is followed by a slowing down due to the hydrodynamic drag force. Furthermore, experiments show that copepods preferentially jump in their onward direction \cite{Buskey-2002, Buskey-2003}. This leads to assume that a general template for the time evolution of velocity during a jump can be adopted. \\
On the mechanical side, the following additional conjectures are made: iii) We assume that copepods are small enough that their center of mass can be considered a perfect fluid tracer in a flow, except for the time when a jump event takes place. In hydrodynamic terms, this means that copepods are assumed to be rigid, homogeneous, neutrally buoyant and with a size that is of the order of the smallest scale of the flow. Neutral buoyancy and mass homogeneity imply that gravity has no role in producing additional acceleration or torque.\\
iv) Finally, copepods are coupled to the fluid in a one-way fashion, which means that they react and are carried by the flow, but they do not modify it; copepods-copepods interactions are also neglected. \\
Gathering all the above assumptions, the LC equation of motion describing a copepod trajectory, $\bm{\mathrm{x}}(t)$, and its body orientation, $\bm{\mathrm{p}}(t)$, reads:
\vspace{.3cm}
\begin{eqnarray}
\dot{\bm{\mathrm{x}}}(t) &=& \bm{\mathrm{u}}(\bm{\mathrm{x}}(t),t) + \bm{J}(t)\\
\label{xdot}
\dot{\bm{\mathrm{p}}}(t) &=& \tfrac{1}{2}\ \bm{\mathrm{\omega}}(\bm{\mathrm{x}}(t),t) \wedge  \bm{\mathrm{p}}(t)
\label{pdot}
\end{eqnarray}
\vspace{.1cm}\\
where $\bm{\mathrm{u}}(\bm{\mathrm{x}}(t),t)$ and $\bm{\mathrm{\omega}}(\bm{\mathrm{x}}(t),t)$ are respectively  the velocity and the vorticity of the carrying fluid flow at the copepod position and $\bm{J}(t)$ is an added velocity term that describes the jump escape reaction of the copepod, and where $\wedge$ is the vectorial product.
We note that eq. (\ref{pdot}) is an accurate description for the rotation rate only for the case of a spherical body, and it is here adopted for simplicity. The generalised form of this equation, valid for axisymmetric ellipsoidal bodies, known as Jeffery equation \cite{Jeffery-1922} can also be used (see its effect on the LC model in \cite{Ardeshiri-2016}).
The jump term, $\bm{J}(t)$, is a function of time but also depends on a set of parameters, which we will introduce and describe in the following.
In our model copepods normally drift with the carrier fluid however, when they find themselves in the alert regions, \textit{i.e.} regions with strain rate larger than the reference threshold value $ \dot{\gamma}_T$, they perform a jump with exponentially decaying velocity intensity over the time. Notice that a jump has a duration $\tau_w = t_e - t_i$, with  $t_i$ and $t_e$ the initial and final time for the jump respectively. Such a time interval can be chosen as the time after which the jump velocity amplitude has declined to a very low value. Note that in our model a new jump cannot occur if the previous jump is not yet finished, this means that the maximal jump rate is  $\tau_w^{-1}$ (see Fig.\ref{jump}). The jumping event  can thus be written as the following expression. When copepods are in alert regions ($\dot{\gamma}(\bm{\mathrm{x}}(t_i),t_i) > \dot{\gamma}_T$):
\begin{equation}
    \bm{J}(t,t_i,t_e,\bm{\mathrm{p}})= 
\begin{cases}
    u_{J}\, e^{\frac{t_i-t}{\tau_{J}}}\, \bm{\mathrm{p}}(t_i), & \text{if } t\leq t_e\\
    0,              & t > t_e
\end{cases}
\label{eq:jump}
\end{equation}
where  $u_{J}$ and $\tau_{J}$ are two parameters characterising the jump shape in terms of, respectively, its velocity amplitude and duration (exponential decay time). The subscript capital $J$ refers to the jump related parameters. 
Note also that, in the above expression, the jump orientation is identified by $\bm{\mathrm{p}}(t_i)$ which is the copepod orientation at the time of the jump activation, $t_i$. The evolution of this orientation is given by eq. (\ref{pdot}).\\ 
According to the experimental measurements on \textit{Eurytemora affinis} and \textit{Acartia tonsa} \cite{Ardeshiri-thesis-2016} a realistic choice for the parameters in the model is $u_J \simeq 10\, cm/s$ and $\tau_J \simeq 10 \,ms$, respectively.  On the contrary our experiments did not allow to estimate the intensity of $\dot{\gamma}_T$ and if any minimal time-interval $\tau_w$ between the jumps exists. In the literature we find that $\dot{\gamma}_T$ for \textit{Acartia tonsa} can be around $0.4 s^{-1}$ \cite{Kiorboe-1999}, however there is a wide range of variability for other species and values as low as $\dot\gamma_{T} = 0.025 \,s^{-1}$ have been also reported for \textit{Acartia tonsa} \cite{Woodson-2005, Woodson-2007}. We add here that, if one assumes valid the exponential model for the jump (\ref{eq:jump}), the time needed to reduce the jump amplitude by a factor $x$ is $\tau_w = \ln(x) \ \tau_J$. Taking then $x=100$ gives approximately $\tau_w \simeq 50 \, ms$. To account for the uncertainties within the parameter values, we conduct a sensitivity analysis by independently varying parameters, $\dot{\gamma}_T$ and $\tau_w$.

\subsection{Fluid flow model}
\label{Flow field simulation}
At small scales ($<\ 1m$) the oceanic turbulence does not depend on large-scale currents, and can be taken as statistically homogeneous and isotropic. In such a condition only one parameter suffices to characterise the intensity of turbulence, namely the Taylor-Reynolds number ($R_{\lambda}$). $R_{\lambda}$ in the ocean can vary in space and time, its most frequent order of magnitude is $R_{\lambda} \simeq 100$. Typical reference values for the properties of such a turbulent flow are given in table \ref{properties-ocean}. 

\begin{table}[htp]
\begin{center}
\begin{tabular}{| c | c | c |}
\hline
Variable & Unit & Value \\
\hline
Kinematic viscosity ($\nu$) & $m^{2}s^{-1}$ & $10^{-6}$ \\
\hline
Mean velocity fluctuation ($u'$) & $m s^{-1}$ & $5\times10^{-3}$ \\
\hline
Turbulent energy dissipation rate ($\epsilon$) & $m^{2}s^{-3}$ & $10^{-6}$ \\
\hline
Kolmogorov length scale ($\eta$) & $m$ & $10^{-3}$\\
\hline
Kolmogorov time scale ($\tau_{\eta}$) & $s$ & 1 \\
\hline
Kolmogorov velocity scale ($u_{\eta}$) & $m s^{-1}$ & $10^{-3}$ \\
\hline
\end{tabular}
\end{center}
\caption{Order of magnitude estimate for the properties of ocean flow at the Taylor Reynolds number $R_\lambda \simeq 100$.}
\label{properties-ocean}
\end{table}%
We adopt as governing equations of the flow the Navier-Stokes equations for an incompressible flow in three-dimensions, which we solve by means of a Direct Numerical Simulation based on a pseudo-spectral method. The fluid domain is a cubic periodic box. The forcing to sustain the flow acts only at large scale leading to a statistically homogeneous and isotropic turbulence at $R_\lambda \simeq 80$.
The forthcoming discussion on the model results is more conveniently addressed in dimensionless units. The reference scales adopted here will be the dissipative (or Kolmogorov) units $\tau_{\eta}$ for time, $\eta$ for space and $u_{\eta} = \eta/\tau_{\eta}$ for the velocity.
The scales are linked by the relation $u_{\eta} \eta / \nu = 1$, and represent the smallest scales of the turbulent flow, \textit{i.e.,} the scales below which the flow can be considered as laminar and dominated by viscous forces (see table \ref{properties-ocean}). The copepods' jump intensity in our simulation turns out to be $u_J/u_\eta = O(10^2)$, while the copepods' jump decay time is $\tau_J/\tau_\eta \simeq O(10^{-2})$. The threshold strain rate thus becomes $\tau_{\eta} \dot{\gamma}_T$.
When the results are analysed in such units, it is expected that the effect of varying $R_\lambda$ (increasing the turbulence intensity) will not change significantly the observed phenomena. This is due to the fact that the LC dynamics, its behavioural reaction, but also its patchiness is linked to the dissipative-scales of the turbulent flow \cite{Ardeshiri-2016}. 
We remark that the above estimate also illustrates the peculiarity of this swimming strategy of copepods. They can impart  a velocity that is even larger than the one of the largest vortices in the flow, larger than $u'$, and this happens over a time which is much shorter than any time scale of the flow. It is clearly an effective way to escape from unwanted locations in the flow.
The jump parameter values, based on our estimates of $R_\lambda$, are $u_J/u_\eta = 250$, and $\tau_J/\tau_\eta = 10^{-2}$. These values remain fixed throughout the study. \\
 
\subsection{Encounter rate in case of preferential concentration}
\label{Encounter rate in case of preferential concentration}
We look at the single-copepod encounter rate, which is the number of encounters per unit time experienced by one copepod in a population of $N$ individuals in  a volume $V$ (number density $n = N/V$).  Under the condition of statistically homogeneous and isotropic movement, the  single-copepod encounter rate can be written in the following form \cite{Saffman-1956, Sundaram-1977,Wang-1998, Collins-2004, Reade-2000, Onishi-2014}:
\begin{equation}
E(r) = n\  2 \pi r^2 \ g(r)\  \langle \delta v_{rad} (r) \rangle  
\label{contact-rate-kernel}
\end{equation}
where $r$ is twice the organism's encounter, or perceptive, radius ($r=2\ r_p$). Notice that here $r_p$ and similarly $r$ can vary from a very small value ($\eta$) to a very large value depending on the organisms size, giving us the possibility to evaluate the encounter rate at different distances. In the end, we will quantify the copepods' encounter rate as a function of their perceptive radius. Here $g(r)$ represents the pair distribution function, which describes the variation of the particles' density from a reference particle: $g(r)=1$ for homogeneously distributed organisms, while $g(r)>1$ in the case of local accumulation. 
Furthermore, for vanishing values of $r$, $g(r)$ is linked to the fractal dimension of a set of points in space, \textit{i.e.}, the correlation dimension $D_2$ \cite{Grassberger-1983}. The correlation dimension is a measure of the dimensionality of the space occupied by a set of random points. Such a dimension can be fractal, \textit{i.e.} to have non-integer value. For a set of aligned evenly spaced points $D_2$ has value 1, a set of points evenly distributed on a surface has correlation dimension 2, while an ensemble of points homogeneously filling a volume has dimension 3. The pair distribution function has the following form:
\begin{equation}
g(r) = \frac{1}{4\pi r^2 n N}\sum_{i=1}^{N}\sum_{j\ne i}^{N} \delta \left(r- | \bm{r}_{ij} |  \right)
\end{equation}
where  $| \bm{r}_{ij} | = \left|\bm{x}_j-\bm{x}_i\right|$ is the Eulerian distance between two particles  in a pair (denoted with indexes $i$ and $j$). The symbol $\delta$ denotes the Dirac generalized function,  with $\delta(0)=1$ and $\delta(x)=0$ for $x \ne 0$.
Finally, $\langle \delta v_{rad} (r) \rangle$ is the mean radial velocity between two organisms separated by the distance $r$:
 \begin{equation}
 \langle \delta v_{rad} (r) \rangle= \frac{ \sum_{i=1}^{N}\sum_{j\ne i}^{N}   \left| \dot{\bm{r}}_{ij} \cdot \frac{\bm{r}_{ij}}{| \bm{r}_{ij} | } \right|  \delta \left(r- | \bm{r}_{ij} | \right) }{ \sum_{i=1}^{N}\sum_{j\ne i}^{N} \delta \left(r- | \bm{r}_{ij} |  \right) }
 \end{equation}
Note that $\dot{\bm{r}}_{ij} = \dot{\bm{x}}_j-\dot{\bm{x}}_i$ denotes here the organisms's velocity difference. A detailed explanation of the appropriate choice of the velocity difference to be used in expression (\ref{contact-rate-kernel}) can be found in \cite{Wang-1998}.
The total encounter rate, which is the total number of encounters in the population is given by $n\ E(r)$.
\\
For swimming microorganisms, such as small algae (e.g. \textit{Cchlamydomonas}) which are neutrally buoyant, a preferential concentration effect has been found resulting from the gyrotactic motility \cite{Durham-2013, DeLillo-2014}, a competition between the spatial gradients in the fluid velocity, that contributes to the vorticity, and the stabilizing torque due to the displacement of the center of gravity from the center of geometry.\\
Compared to phytoplankton, copepods show a different type of complexity due to their reactive behavior. Contrary to the previously observed clustering, the patchiness of copepods is tightly linked to their behavioural strategy in turbulent flows; this is the central result in our previous study of the LC model \cite{Ardeshiri-2016}.\\

\section{Results}
\label{Results}
We organize the analysis in two sections. The first addresses the dependence of encounter rates on the shear rate threshold $\dot{\gamma}_T$ (in dimensionless unit, on $\tau_{\eta} \dot{\gamma}_T$) at keeping constant the maximal inter-jump frequency $\tau_w^{-1}$. The second section instead fixes $\dot{\gamma}_T$ and allows $\tau_w$ to vary. 
\subsection{EFFECT OF THE DEFORMATION RATE THRESHOLD}
\label{Effect of the shear rate threshold}
\begin{figure}[htpb]
  \centering
             \includegraphics[scale = 0.33, angle=-90]{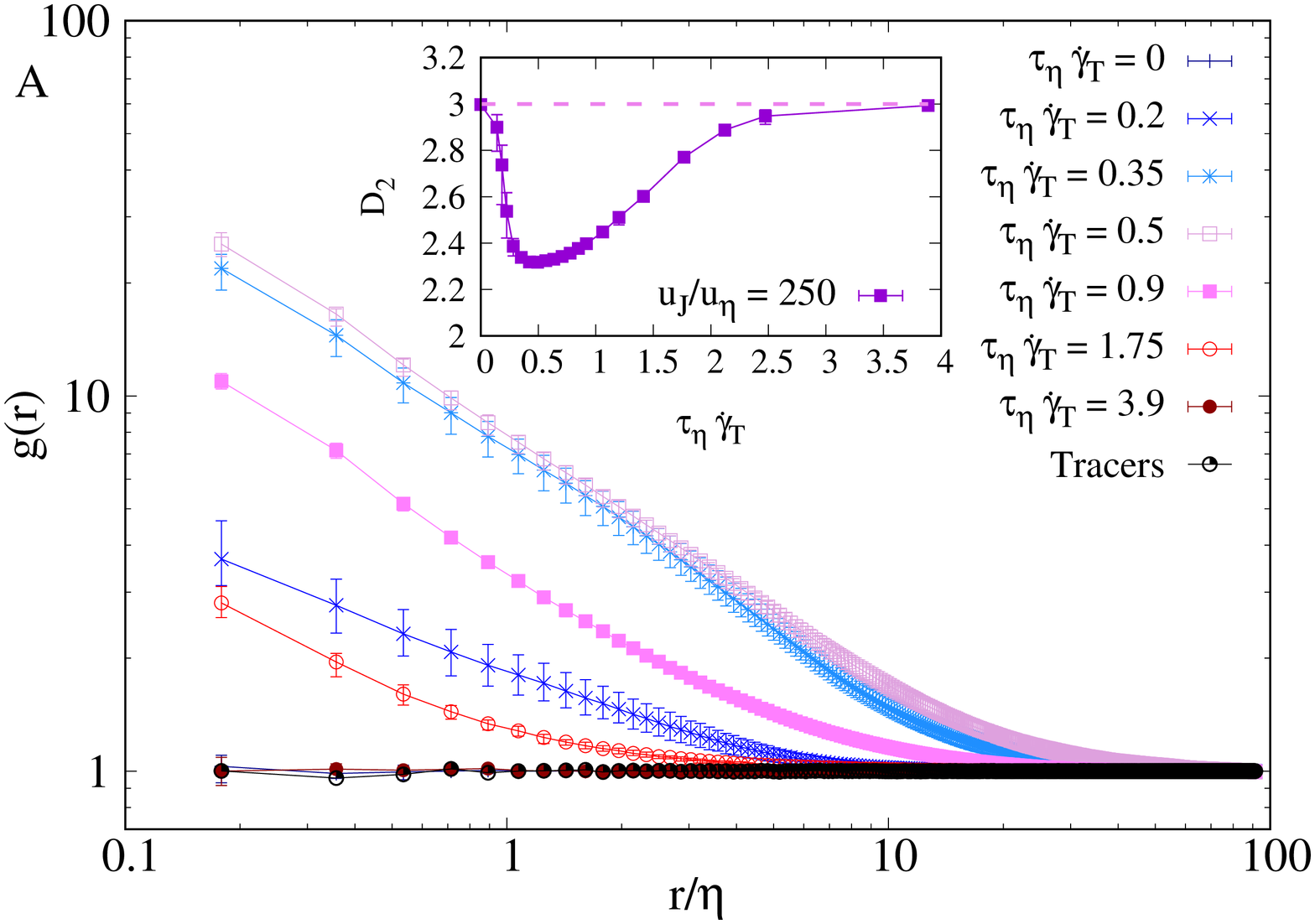}
             \includegraphics[scale = 0.33, angle=-90]{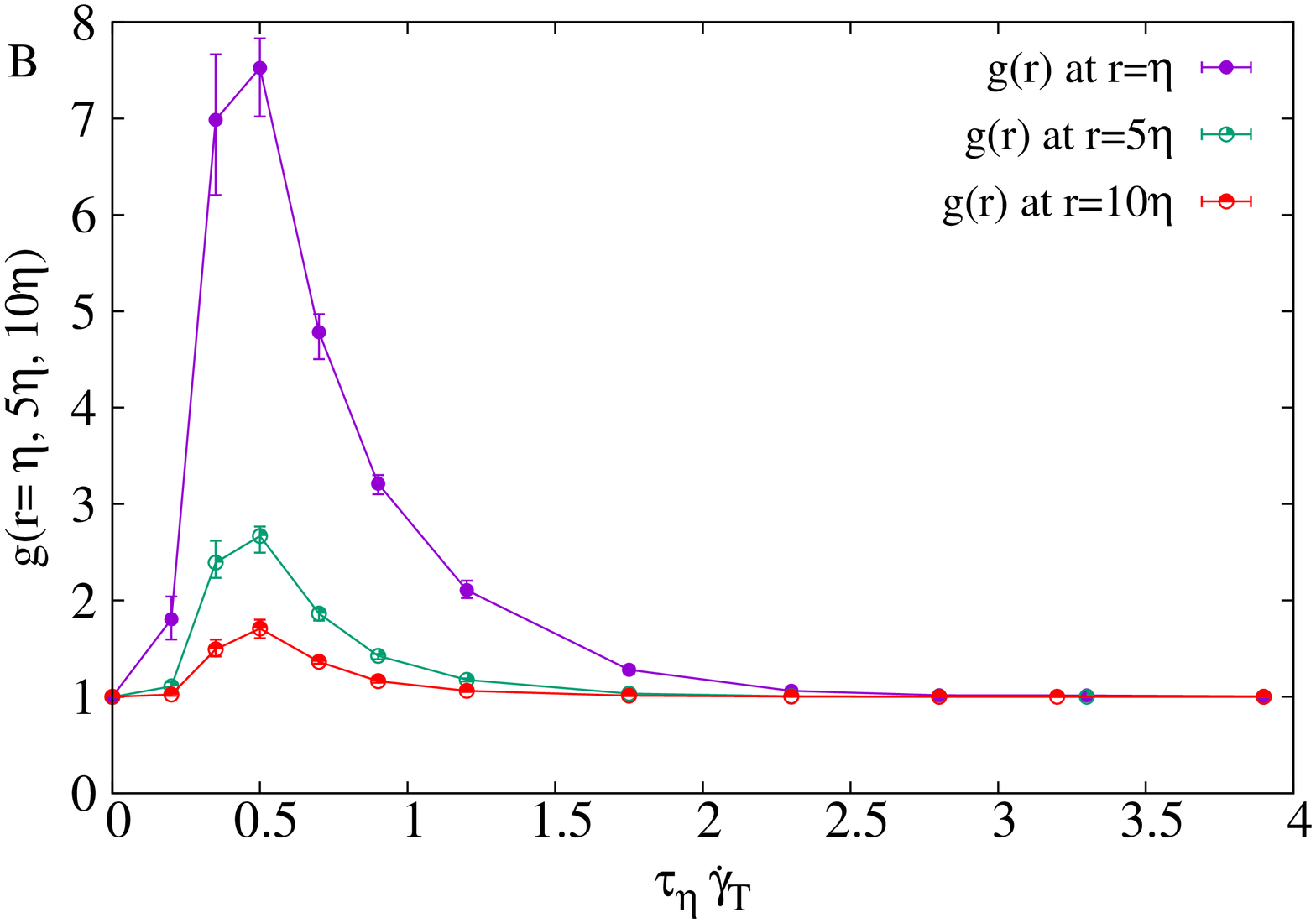}
 \caption{(A) Pair-radial-distribution function, $g(r)$, for different Lagrangian copepod families with different threshold values of the dimensionless shear rate $\tau_\eta \dot{\gamma}_T$. The inset represents the correlation dimension of copepod distribution with jump intensity, $u_J/u_\eta = 250$ and $\tau_J/\tau_\eta = 10^{-2}$. (B) Variation of $g(r)$ at different distances as a function of the threshold $ \tau_\eta \dot{\gamma}_T$.}
 \label{pair-correlation-D2}
\end{figure}

The pair correlation function for the LC model, one of the two main factors in the encounter rate expression (\ref{contact-rate-kernel}),  is shown in Figure \ref{pair-correlation-D2} (A). The presence of local copepods concentration is here evident from the fact that $g(r) \gg 1$ at small values of $r$. Its trend however is non-monotonic with the shear rate threshold parameter. It reaches a peak value (here of about ~30) for $\tau_\eta \dot{\gamma}_T = 0.5$, then it decreases. 
The same trend is observed for the steepness of the decreasing slope of $g(r)$. This reflects what was previously observed \cite{Ardeshiri-2016} for the correlation dimension $D_2$  (see the inset of Fig. \ref{pair-correlation-D2}A).
Indeed it is known that the following relation applies $$\lim_{r\to 0} g(r) \sim r^{D_2-3}$$ where 3 stands for the dimension of the physical space. 
The inset of Fig. \ref{pair-correlation-D2}A reports the $D_2(\tau_{\eta} \dot{\gamma}_T)$ dependence. We note the value at minimum $D_2 \sim 2.3$ for $\tau_\eta \dot{\gamma}_T = 0.5$, which suggests that copepods may form almost bidimensional, sheet-like structures, like non-planar layers \cite{Ardeshiri-2016}.\\
Fig. \ref{pair-correlation-D2}B is another way to consider the same measurement: it represents the intersection with vertical cuts of Fig. \ref{pair-correlation-D2}A with value $r/\eta = 1,5$ and $10$. For each curve there is a maximum for $\tau_{\eta} \dot{\gamma}_T=0.5$ and this maximum is much larger at $r/\eta = 1$. 
It is here important to note that the range of $\tau_{\eta} \dot{\gamma}_T$ values, in which the preferential concentration arises is very narrow.  $\tau_{\eta} \dot{\gamma}_T$ outside the range $[0.125,2]$ leads to almost negligible clustering (see Fig. \ref{pair-correlation-D2}B). Furthermore we see clearly that the larger the perception radius, the less effective is the preferential clustering mechanism. We may observe that for $r>10\eta$, which means perception radius larger than $5 \eta$ we can take $g(r) \simeq 1$ and hence there is no more clustering effect in the encounter rate.

\begin{figure}
  \centering
             \includegraphics[scale = 0.33, angle=-90]{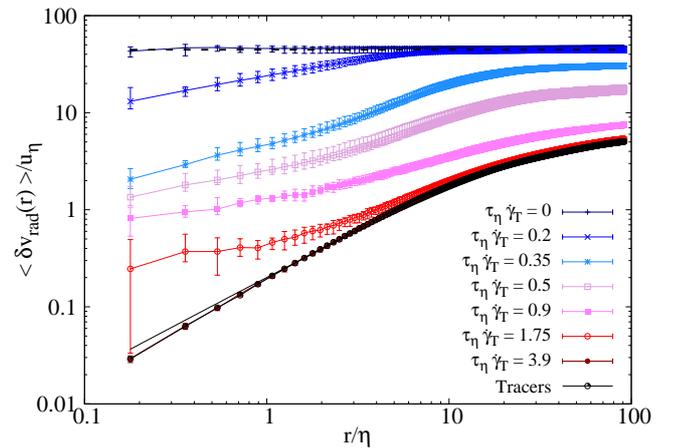}
  \caption{Dimensionless mean radial velocity, $\langle \delta v_{rad} (r) \rangle / u_\eta$ between couples of Lagrangian copepods separated by a distance $r/\eta$ for different values of the dimensionless threshold strain rate intensity $\tau_\eta \dot{\gamma}_T$. The continuous line is the prediction for fluid tracers based on the empirical approximation on the second-order longitudinal Eulerian structure function proposed by Borgas and Yeung \cite{Borgas-2004}. The dashed line plateau indicates the prediction derived from a random field of spatially and temporally uncorrelated jumps.}
  \label{Fig2-Wr}
\end{figure}

\begin{figure}
  \centering
  \includegraphics[clip,trim={2.cm 4cm 2.5cm 0},width=6.5cm, angle=-90]{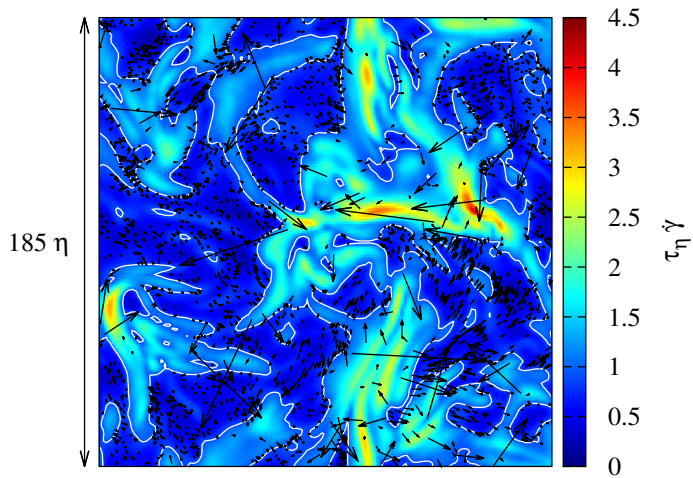}
  \caption{Visualisation of the instantaneous spatial distribution of Lagrangian copepods in turbulent flow along with their velocity vectors. The data here comes from a two-dimensional slice, of spatial lateral dimensions $185 \eta \times 185 \eta$ and thickness $\eta$, out of the simulated $(185 \eta)^3$ three-dimensional domain. The velocity vectors are in a arbitrary scale. 
The copepods are characterised by the parameter set: $u_J= 250 u_{\eta}$, $\tau_J = 10^{-2} \tau_{\eta}$, $\dot{\gamma}_T = 0.91\ \tau_\eta^{-1} $ and $\tau_w = 4.6 \tau_J$. 
  The color map represents the instantaneous Eulerian field of the strain rate, $\left| \dot\gamma \right|$. Contour lines are traced for the threshold value  $ \dot\gamma = \dot\gamma_T $, hence it traces the boundaries between comfort and alert regions in the flow (respectively light and dark shaded regions).}
  \label{output115}
\end{figure}

\begin{figure}
  \centering
             \includegraphics[scale = 0.33, angle=-90]{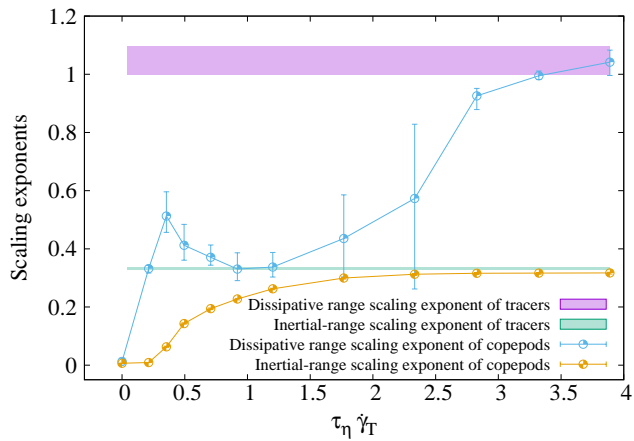}
  \caption{Scaling exponents of $\langle \delta v_{rad} (r) \rangle$ vs. $r$ from power law fits in the dissipative ($r/\eta$ in range [0.15,2.5]), $r^{\zeta_d}$, and inertial-range limit ($r/\eta$ in range [35,100]), $r^{\zeta_i}$. Error bars on exponents have been estimated by shifting the fitting window towards lower and higher value.}
  \label{Fig5-scaling-exponent}
\end{figure}

The mean radial velocity between two copepods at varying $\tau_{\eta} \dot{\gamma}_T$ is shown in Fig. \ref{Fig2-Wr}. For comparison the same quantity for fluid tracers is also shown. We note that when computed on fluid tracers the radial structure function is an Eulerian quantity which, under the hypothesis of isotropic turbulence, can be related to the better known second-order longitudinal Eulerian structure function via the relation  $\langle \delta v_{rad} (r) \rangle = \sqrt{ 2 \langle \delta v_{\parallel}^2 (r) \rangle / \pi }$ \cite{Zaichik-2006}.
If we adopt the empirical approximation given by Borgas and Yeung \cite{Borgas-2004} for $\langle \delta v_{\parallel}^2 (r) \rangle$ at finite $R_\lambda$, we find a pretty good agreement with our numerical results. 
In turbulence, the Eulerian mean radial velocity grows linearly as $r^{\zeta}$ with $\zeta = 1$ for dissipative scales ($r \lesssim 10 \eta $) and  $r^{\zeta}$ with $\zeta = 1/3$ at inertial-range scales \cite{Frisch-book}. From Fig. \ref{Fig2-Wr} we see that the jump-rate of copepods with threshold value of $\tau_\eta \dot{\gamma}_T \geq 3.9$ is so low that they behave almost like tracers. They are thus passively advected by the flow and, for this reason they go very close to the prediction just given for tracers. By decreasing the shear rate threshold value, copepods become more and more reactive, therefore the jumping part in the mean radial velocity expression becomes increasingly dominant.
In the case where all copepods are permanently in alert regions ($\tau_\eta \dot{\gamma}_T =  0$) one can expect a Brownian-like motion of copepods according to the LC model.  For this case, copepod's relative velocity tends to be constant over space and its amplitude is proportional to the jump intensity ($u_J$).  If we assume that the jumps for different copepods are uncorrelated both in time and in space and that the additional fluid velocity is negligible one can easily compute the level value of this plateau (see Fig. \ref{Fig2-Wr}).
By dimensional reasoning one can show that in the latter case the effective diffusivity of the copepods is proportional $u_J^2\tau_J$. For other cases (intermediate shear rate threshold values $\tau_\eta \dot\gamma$) the behaviour is more complicated and difficult to capture by analytical or dimensional arguments. 
We observe that in the limit of $r\to 0$ the mean radial velocity function, $\langle \delta v_{rad} (r) \rangle$, goes to zero for fluid tracers but it has non-zero value for different families of copepods and is more pronounced by decreasing the shear rate threshold value. This pattern is caused by singularities in the copepods dynamics which implies that copepods at distance $r$ may have a different behaviour, hence a different velocity. In the field of particle laden flows this discontinuity in the particle velocity field is often referred to as a caustic singularity \cite{Crisanti-1992, Gustavsson-2012}.  This is better illustrated in the visualisation of Fig. \ref{output115} where pairs of very close copepods may have very large velocity differences.\\ 
Finally, we remark  that differently from the trend observed for $g(r)$ the behaviour of $\langle \delta v_{rad} (r) \rangle$ is inversely proportional and monotonic with $\tau_\eta \dot{\gamma}_T$ : at increasing the shear rate threshold the mean radial velocity difference goes down from the plateau level to the fluid tracer level.
The scaling exponents $\zeta$, of $\langle \delta v_{rad} (r) \rangle$ for different copepod families at varying $\tau_\eta \dot{\gamma}_T$ via power law fits are shown in Fig. \ref{Fig5-scaling-exponent}. This is to see how copepods' dynamics change compared to the fluid tracers, as a function of the threshold strain rate. The change of the power law scaling exponent in the inertial-range is smooth, but at the dissipative scale it shows a non monotonic behaviour for the fractal dimension as a function of the strain rate threshold values, as already reported for the pair correlation function in Fig. \ref{pair-correlation-D2}.

\begin{figure}
  \centering
           \includegraphics[scale = 0.33, angle=-90]{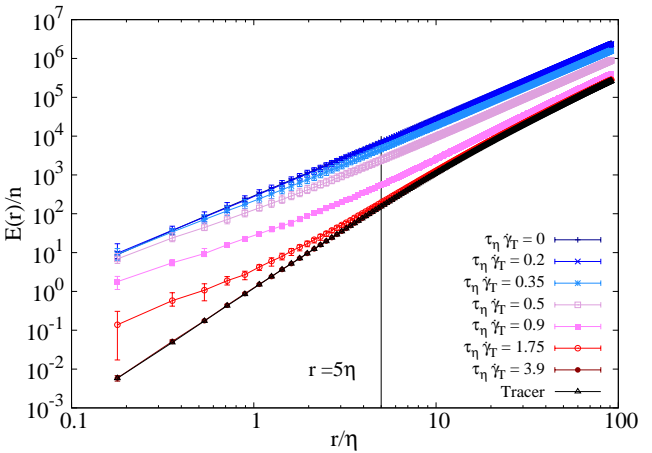}
  \caption{Encounter rate per unit particle density for different Lagrangian copepod families with different values of deformation-rate threshold $\tau_\eta \dot{\gamma}_T$.}
  \label{Fig3-Kernel}
\end{figure}

Having the pair correlation function $g(r)$, and the mean radial velocity $\delta v_{rad}(r)$, one can estimate the encounter rate kernel $E(r)$. There is of course no contact between fluid tracers, because they are size-less by definition and are passively advected by an incompressible flow, where the fluid streamlines does not cross each other. However, when a virtual radial size $R$ is assigned to a tracer, their virtual contact rate is proportional to the mean radial velocity evaluated at $r=2 R$.
Copepods differ from tracers in two aspects, first they are not simply passively transported by the flow and secondly their virtual contact rate, called radius of perception,  has a specific biological meaning. It is  known that copepods-copepods or copepods-prey/food interactions occurs before the possible body-to-body physical contact: copepods can grab their prey or become aware of an incoming mate before hit them \cite{Schmitt-2008}. This is why $R$  should be understood as the radius of perception of the organisms rather than their geometrical size.\\
Fig. \ref{Fig3-Kernel} reports the encounter kernel as a function of $r$ for different copepod families. Here the encounter rate of tracers are depicted as for comparison with the different copepod families.  It appears that by increasing the strain rate threshold value ($\tau_\eta \dot{\gamma}_T$) the encounter rate of copepods decreases monotonically, although the growth of the shear rate $\tau_\eta \dot{\gamma}_T$ had non-uniform impact on the pair correlation dimension (see Fig. \ref{pair-correlation-D2}). This implies that the dominant term in interspecies encounter rate of the copepods is the amplitude of their mean radial velocity. The encounter rate is dominated by caustics.

\begin{figure}
  \centering
             \includegraphics[scale = 0.33, angle=-90]{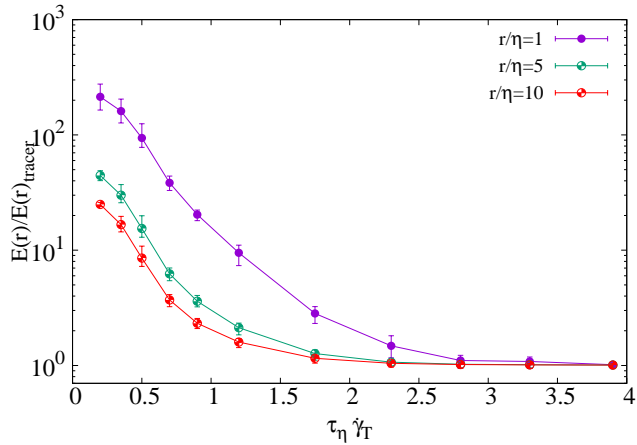}
  \caption{Ratio between encounter rates experienced by Lagrangian copepod particles and the one experienced by fluid tracer particles with the same perception radius.}
  \label{Fig4-Kernel-comparison-log}
\end{figure}

The vertical line in the figure shows the radius of perception for Lagrangian copepod particles, here supposed to be $5$ times greater than the Kolmogorov length scale of the carrier fluid. The ratio of the radius of perception to the copepods' body size is reported to be in the range $1-3$ \cite{Lenz-1993,Bagoien-2005, Doall-1998}. In terms of $\eta$ this means that the radius of perception of copepods is of the order of $\sim 1$. In order to see how effective the deformation rate threshold value is on the encounter rate of Lagrangian copepods at different perception radius, one can estimate the ratio between the encounter rates experienced by copepod families and tracers. This is shown in Fig. \ref{Fig4-Kernel-comparison-log} where it is realistic to have larger encounter rates at small distances. This figure suggests that at optimum clustering, corresponding to the shear rate threshold value of $\tau_\eta \dot{\gamma}_T = 0.5$, the encounter rate can be of the order of $\sim 10$ with respect to the tracers at distance $r = 5\eta$. The LC model shows no contact rate enhancement at shear rate values larger than $2.75$. This means that when $\dot{\gamma} > 2.75/\tau_\eta$ then there is any increase of contact rate between copepods due to their swimming behaviour. Therefore, dimensionally the LC model do not give any significant advantage at $\dot{\gamma} > 2.75\,s^{-1}$. When the shear rate $\dot\gamma$ is lowered, the contact rate reaches up to $100$ times the value of fluid tracers.  
Notice however that copepod-copepod interaction has been neglected in the LC model, and this may have lead to large overestimation of contact-rates.
\subsection{Effect of waiting time between successive jumps}
\label{Effect of waiting time between successive jumps}
\begin{figure}
  \centering
             \includegraphics[scale = 0.35, angle=-90]{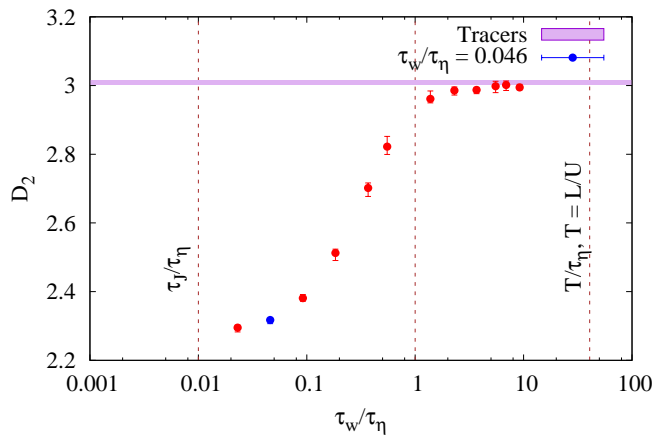}
  \caption{Effect of the waiting time between successive jumps on the correlation dimension. Blue dot corresponds to the copepods family with $u_J/u_\eta = 250$, $\tau_J/\tau_\eta = 10^{-2}$, $\dot{\gamma}_T = 0.5 \tau_\eta^{-1}$ and $\tau_w/\tau_\eta = 0.046$ for which optimal clustering happened. The vertical dashed lines show the $\tau_J/\tau_\eta$, $\tau_w = \tau_\eta$ and the ratio between the large eddy turnover time of the flow, $T$, and Kolmogorov time scale,$\tau_\eta$, from left to right. 
  Note that in this test $\dot{\gamma}_T = 0.5 \tau_\eta^{-1}$ is kept fixed and corresponds to the case with maximum small scale clustering.}
  \label{Inaction}
\end{figure}

Up to now we assumed that a copepod's jump was terminated after a prescribed time interval $\tau_w$ from the jump inception, defined it as when the jump velocity amplitude reaches one percent of the copepods' initial jump intensity. In reality copepods can behave differently. For instance they can be less reactive of what has been modeled so far since they have some finite  energy amount available for swimming and after repeated jumps they may need some time in order to recover their lost energy \cite{Visser-2003}. The probability density function of time between successive jumps for \textit{E. Affinis} and \textit{A.Tonsa} (not shown here) indicates the presence of memory on the previous jumps of copepods (see also \cite{Dur-2010}) but there is a lack of quantitative elements in order to have enough information to model this feature in the LC model.\\
We now vary the duration of the waiting time, taking as a reference the family for which $\tau_w/\tau_J = c$ with $c=4.6$ (a family with $u_J/u_\eta = 250$, $\tau_J/\tau_\eta = 10^{-2}$ and $\tau_\eta \dot{\gamma}_T = 0.5$ where a complete jump lasts for $46\,ms$) as shown in blue in Fig.\ref{Inaction}. We now have Lagrangian copepod families with $u_J/u_\eta = 250$, $\tau_J/\tau_\eta = 10^{-2}$, $\tau_\eta \dot{\gamma}_T = 0.5$ and $\tau_w/\tau_J$ which varies in range $[0.5c, 200c]$. 

Fig. \ref{Inaction} interestingly shows that by making copepods progressively less reactive, the small scale clustering, characterised by $D_2$, fades away and only a prompt copepod reactions may lead to non-homogenous spatial distribution. Moreover, when the waiting time, $\tau_w$, is larger than the Kolmogorov time scale of the flow, $\tau_\eta$, copepods' spatial distribution becomes nearly homogenous. 

\begin{figure}
  \centering
             \includegraphics[scale = 0.33, angle=-90]{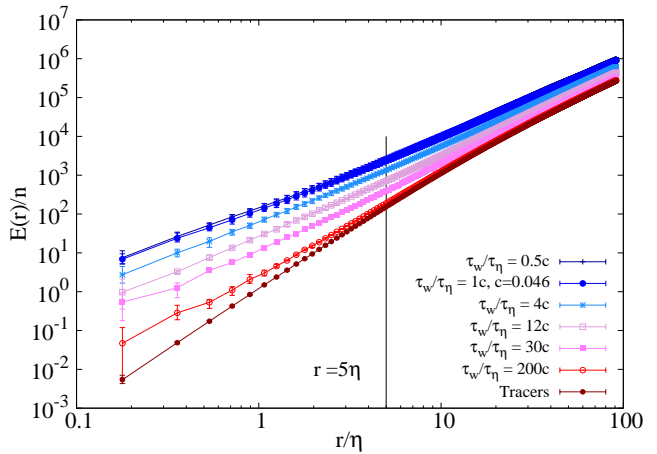}
  \caption{Encounter rate per unit particle density for different Lagrangian copepod families with different waiting time between successive jumps, $\tau_w/\tau_\eta$ (all other parameters are the same as in previous figure).}
  \label{Fig3-Kernel-jumprest}
\end{figure}

\begin{figure}
  \centering
             \includegraphics[scale = 0.33, angle=-90]{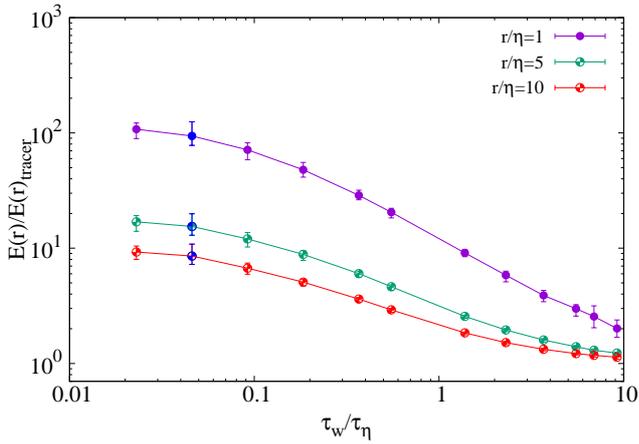}
  \caption{Ratio between encounter rates experienced by Lagrangian copepod particles and the one experienced by fluid tracer particles with the same perception radius as a function of waiting time between successive jumps. The filled symbols correspond to the copepods family with $\tau_w/\tau_J = c$ at different perception radius.}
  \label{Fig4-Kernel-comparison-log-jumprest}
\end{figure}

It is now interesting to see how these changes in spatial distribution can affect the encounter rate. In Fig. \ref{Fig3-Kernel-jumprest} the encounter rate per unit particle density as a function of $r$ is shown for different Lagrangian copepod families that we observed in Fig. \ref{Inaction}.
One can normalize the encounter rate experienced by copepod families by the one of tracers at a specific perception radius in order to clearly see the effect of waiting time between successive jumps. This is shown in Fig. \ref{Fig4-Kernel-comparison-log-jumprest} where it is seen that the increase of the waiting time decreases the copepods' encounter rates.  For $\tau_w \simeq 2 \tau_{\eta}$ the encounter rate enhancement as compared to tracers is just a factor $\sim 8$ for $r=\eta$ and decrease to nearly a factor $\sim 2$ for $r=5\eta$.

\section{Discussion}
\label{Discussion}
The enhancement of the encounter rate between individuals from the same species was achieved here by using the proposed Lagrangian copepod model in turbulence. It was observed that this enhancement does not occur necessarily at copepods' optimum preferential concentration. This increase has its origin in two distinct mechanisms: the preferential concentration, quantified by $g(r)$, and and the mean radial velocity, measured by $\delta v_{rad}(r)$.
The parametric analysis at varying $\tau_\eta \dot{\gamma}_T$ reveals that the lower the threshold, the higher the encounter rate is. A similar result is found for the investigation at fixed $\dot{\gamma}_T  = 0.5 \tau_{\eta}^{-1}$ and at varying inter-jump minimal time $\tau_w$: the shorter the inter-jump time, the more frequent the encounters are, implying $\delta v_{rad}(r)$ to be the dominant term in encounter rate formulation. These results are easy to grasp. A very reactive organism will explore much more space than a nearly passive one and will enhance its chances to meet similar organisms. However, one may want to ask if, despite its relative simplicity, the LC model can provide also a nontrivial insight in the possible dynamics of copepods in a turbulent flow.\\
A first significant observation is that the dissipative scale, $\tau_{\eta}$ is the relevant control scale for the encounter rate enhancement. It defines respectively both the reference frequency and the reference time gap for $\dot{\gamma}_T$ and $\tau_w$.
In other words the present mechanisms of enhancement of encounters is effective only if copepods have a shear rate sensitivity that is finer than the shear rate produced at the Kolmogorov scale $\sim \tau_{\eta}^{-1}$ and if their jumps occur at a rate higher than this very same frequency. In dimensional terms this corresponds to $\tau_{\eta}^{-1} \simeq 1 s^{-1}$, however, due to the difference turbulent conditions found in the ocean, it is known it might have more than one order of magnitude of variability (from 0.1 to 10 $s^{-1}$ as from \cite{Jimenez-1997}) . The fact that the shear rate threshold value to trigger a jump  in copepods has been reported to vary from $\dot\gamma_{T} = 0.025 \,s^{-1}$ \cite{Woodson-2005, Woodson-2007} to $\dot\gamma_{T} = 0.4 \,s^{-1}$ \cite{Kiorboe-1999}, seems to support the validity of the model findings. 
A second observation is that the enhancement  of contact rate by the present mechanism is much less effective when the perception radius is wider. We note the large gap existing between the estimated encounter rates at $r=\eta$ and $r=10 \eta$. This has a biological relevance: for larger copepods or copepods simply with a larger perception range, the preferential concentration mechanisms and the mean radial velocity have less effect on the contact rate. This also sets a quantitative limit, if $r \geq 10 \eta \simeq 1 cm $ the effect of encounter rate enhancement by turbulence is negligible. The fact that the radius of perception for copepods is in the $mm$ range  \cite{Lenz-1993,Bagoien-2005, Doall-1998} suggests that the proposed mechanisms may be effective at least for small copepod species. A complementary interpretation of this result is that for a copepod family of a given perception radius, the turbulence induced enhancement of encounter rates is less and less effective as the the turbulent intensity is increased, and is more effective at moderate levels of turbulence.\\
We now discuss possible limitations of the model and potential features that we neglected and that may change the scenario described so far. 
One shortcoming of the model is the fact that we do not take into account the energetics of copepods. It is unlikely that these organisms may jump indefinitely at a  maximum rate $\tau_w^{-1}$, even if such a rate is not sustained. It is more likely that periods of high activity will be followed by a resting time. A realistic energetic model for copepods could lead to slightly different quantitative estimates for the encounter rate kernel $E(r)$, however it will not modify the observation that $\tau_{\eta}$ is a characteristic scale of the problem. Another limitation concerns the absence of a post-encounter dynamics in our model. The Lagrangian copepods we simulate can have trajectories that cross each others and do not bounce or react differently when they encounter each other (we talk in this case of virtual encounters). In reality we expect different behaviour in the moments following an encounter. This may lead to substantial different conclusions and we think it will be interesting to include this feature as a refinement of the present model. \\
The definition or the estimation of the encounter rate between planktonic organisms is a fundamental concept from the ecological point of view. In fact, many crucial processes occurring at small scales (\textit{i.e.} mating, predator-prey interactions, etc) and having important implications on the plankton dynamics are dependent on the encounter rate concept. For copepods, it is not easy to extrapolate empirical expressions of encounter rates obtained in still conditions \cite{Michalec-2015b} to realistic situations with well-developed turbulence. Our model can be considered as a first step towards a better quantification of the encounter rates of copepods under realistic fluid motion. Our results confirmed that the interaction between turbulence and copepod jumping is not trivial and should be deeply explored from the experimental point of view. In fact, only few studies focused on the jumping behaviour in turbulent conditions. The recent study by Michalec et al., \cite{Michalec-2015b} suggested that the increase in swimming effort (\textit{i.e.} acceleration) when turbulence increase is a kind of compensatory response to the increase in flow velocity. The same authors suggested that the capacity of jumps in copepods, and mainly those inhabiting turbulent areas (\textit{i.e.} estuarine and coastal environments), is a crucial trait to better understand their ecology. It appears from the same experimental study that a threshold value of turbulence can modulate copepod behaviour. In other words, copepods can use their capacity of jumping only when it is useful and certainly not under high turbulent conditions. Our results suggested that the threshold turbulence value could be a species-specific property that should be better estimated in the future using adequate experimental designs. Moreover, our simulations suggested that the copepods' optimum preferential concentration is an important property that should be studied experimentally.

\subsection{CONCLUSION}
\label{Conclusion}
This paper addresses the problem of the quantification of intraspecies contact rate in copepods under realistic conditions. The copepods inhabit flows characterised by large-scale currents and turbulence of variable intensities. The copepods' encounter rate is certainly influenced by this turbulent environment, but on the other hand copepods are also known to be capable to displace quickly when locally subjected to mechanical disturbances.

The way in which environment and this single, specific, behavioural conditions combines is complex. In this paper we couple the exact dynamics of homogeneous and isotropic turbulent flows, by using DNS, with a simplistic model of behaviour where copepods jumps are triggered by localized high strain events. The main result of this investigation lies in the enhanced intraspecies contact rate with respect to the case where copepods are considered as fluid tracer particles. This enhancement comes from two terms in the contact rate expression; one is the variation of the pair correlation function $g(r)$ which accounts for the spatial preferential concentration (patchiness of copepods) and the other one is the variation of the mean radial velocity $\langle \delta v_{rad} (r) \rangle$, which comes from the fact that copepods can have an independent swim velocity.  
Our analysis shows that the encounter rate for copepods of typical perception radius of $\sim \eta)$, where $\eta$ is the dissipative scale of turbulence, can be increased by a factor up to $\sim 10^2$ compared to the one experienced by passively transported fluid tracers, a very large value, which can be ecologically important.  Such effect may show that jumping behaviour of copepods is ecologically justified not only to avoid predators, but also to keep individuals within patches (\textit{i.e.} locally high concentration of individuals) and increasing the encounter rate between congeners that can enhance mating rates. Furthermore, the study of a minimal pause interval between consecutive jumps shows that any encounter-rate enhancement is lost if such time goes beyond the dissipative time-scale of turbulence $\tau_{\eta}$. This provides relevant constraints on the turbulent-driven enhancement of contact-rate due to a purely mechanic induced escape reaction.\\
We conclude by remarking that the large enhancement of the contact rate highlighted in this paper, while relevant for mating behaviour of copepods it is less relevant for prey capture estimates. As a perspective it would be interesting to see the consequence of our LC model for the feeding behaviour by using the Lagrangian model of copepods one side and larger bodies drifting in the flow which can model the presence of large predators.\\
Finally it is important to stress that, in addition to swimming behavior induced by changes in external flow conditions, other mechanisms can also play a role in encounter rates for copepods, \textit{e.g.} chemoreception and mechanoreception  \cite{Buskey-1984, Weissburg-1998}, prey movement detection \cite{Visser-2001, Jiang-2002} and feeding currents \cite{Marrase-1990}. All these potentially relevant effects are outside the scope of the present study: we deliberately targeted the interactions between copepod jumping and turbulence, which received less attention in previous ecological and/or modeling studies.\\

\section*{Funding}
\label{Funding}
This study is part of the PhD thesis of H. Ardeshiri. This work was financially supported by a grant for interdisciplinary research ``Allocation President 2013" of the Universit\'{e} de Lille 1. The authors would like to acknowledge networking support by COST Action MP1305 ``Flowing matter".  E.C. thanks Haitao Xu for useful discussions. 
\bibliography{citation}

\end{document}